\input harvmac.tex
\noblackbox
\lref\stromvafa{A. Strominger and C. Vafa, Harvard and Rutgers 
  preprint HUTP-96/A002 and RU-96-01, January 1996, hep-th/9601029.}
\lref\callanmalda{C. Callan and J. Maldacena, Princeton preprint 
  PUPT-1591, February 1996, hep-th/9602043.}
\lref\horstrom{G. Horowitz and A. Strominger, Santa Barbara preprint,
  February 1995, hep-th/9602051.}  
\lref\polchinski{J. Polchinski, Phys. Rev. Lett. {\bf 75} (1995) 4724 
  [hep-th/9510017].}
\lref\myersperry{R.C. Myers and M.J. Perry, Ann. Phys. {\bf 172} 
  (1986) 304.}
\lref\senreview{See, for example, A. Sen, Int. J. Mod. Phys. {\bf A9} 
  (1994) 3707 [hep-th/9402002]. }
\lref\horsen{G.T. Horowitz and A. Sen, Santa Barbara preprint 
  UCSBTH-95-27, September 1995, hep-th/9509108.}
\lref\nonext{To appear.}
\lref\vafawitten{C. Vafa and E. Witten, Nucl. Phys. {\bf B431} (1994) 
  3 [hep-th/9408074]. }
\lref\vafagas{C. Vafa, Harvard Preprint HUTP-95-A042, November 1995,
   hep-th/9511088.}
\lref\bersadvafa{M. Bershadsky, V. Sadov and C. Vafa, Harvard 
  preprint HUTP-95-A047, November 1995, hep-th/9511222.}
\lref\vafainst{C. Vafa, Harvard preprint HUTP-95-A049, December 1995,
  hep-th/9512078.}
\lref\bouchfriedkent{W. Boucher, D. Friedan, A. Kent, Phys. Lett. 
  {\bf B172} (1986) 316.}
\Title{HUTP-96/A005, McGill/96-07, PUPT-1592}
{\vbox{\centerline{\bf D--branes and Spinning Black Holes}}}
\medskip
\centerline{J.C. Breckenridge\footnote{$^\clubsuit$}
  {e-mail: {\tt jake@haydn.physics.mcgill.ca}},
  R.C. Myers\footnote{$^\diamondsuit$}
  {e-mail: {\tt rcm@hep.physics.mcgill.ca}},     } 
\centerline{\it Physics Department, McGill University,}
\centerline{\it Montreal, PQ, H3A 2T8, Canada}
\smallskip
\centerline{A.W. Peet\footnote{$^\spadesuit$}
  {e-mail: {\tt peet@puhep1.princeton.edu}},}
\centerline{\it Joseph Henry Laboratories, Princeton University}
\centerline{\it Princeton, NJ 08544, USA}
\smallskip
\centerline{and}
\centerline{C. Vafa\footnote{$^\heartsuit$}
  {e-mail: {\tt vafa@string.harvard.edu}}}
\centerline{\it Lyman Laboratory of Physics, Harvard University,}
\centerline{\it Cambridge, MA 02138, USA}
\medskip

\centerline{\bf Abstract}
We obtain a new class of spinning charged extremal black holes in five
dimensions, considered both as classical configurations and in the
Dirichlet(D)--brane representation.  The degeneracy of states is
computed from the D--brane side and the entropy agrees perfectly with
that obtained from the black hole side.

\smallskip

\Date{2/96}
\eject
\newsec{Introduction}

Recently, significant progress has been made in understanding the
degrees of freedom giving rise to the entropy of certain black holes
in string theory \stromvafa\ (and references therein)\foot{See also
the recent works \callanmalda, \horstrom}.  This was achieved by using
the beautiful representation of solitons carrying Ramond--Ramond
charge as D--branes \polchinski.

In this work we report additional progress in this direction, by
investigating spinning black holes in five spacetime dimensions, in
theories with $N=4$ supersymmetry\foot{This easily extends to the case
of Type II compactified on $T^5$, which has $N=8$ supersymmetry.}.
The black holes which we will consider carry electric charge $Q_F$ and
an antisymmetric tensor charge $Q_H$, and are spinning generalizations
of the solutions of \stromvafa.  We concentrate on BPS--saturated
states, i.e. extremal black holes, so that we may rely on adiabatic
arguments for the invariance of the expression for the entropy under
changes in the string coupling.

We construct black hole solutions which have equal--magnitude angular
momenta in the two independent planes, i.e. $|J_1|=|J_2|\equiv J$.
The spinning black hole entropy is computed and is found to be
$$
S_{BH} = 2\pi\sqrt{{{Q_H\,Q_F^2}\over{2}}-J^2}
$$
The entropy obtained from counting spinning D--brane states is
found, for large charges, to be
$$
S_{micro} =  2\pi \sqrt{Q_H({1\over 2}Q_F^2+1)-{1\over 4}
(|J_1|+|J_2|)^2}
$$
This is in exact agreement with the black hole answer in the case of
interest, namely for large charges $Q_{H,F}$ and spins, as well as
$|J_1|\simeq|J_2|=J$.
We find it remarkable that the agreement is precise, including the
crucial numerical factors.

This paper is organized as follows.  In Section 2 we review actions of
heterotic string theory on $T^4$ and Type II theory on $K3$ and
subsequent compactifications to five dimensions on an $S^1$, and
outline our method for generating the solution.  Section 3 contains a
summary of the black hole solutions.  In Section 4, we explain the
counting of states from the D--brane side, and we end with some
comments in Section 5.

\newsec{Actions and solution--generating}

We will begin with a five dimensional black hole which spins in a
single plane, and is a solution of the five--dimensional Einstein
equations.  We add a trivial flat dimension with coordinate $y$, and
the metric is thus \myersperry\
\eqn\kerrfived{\eqalign{
ds_6^{2} 
&= G_{6\,\mu\nu} dx^\mu dx^\nu \cr
&= - dt^2 + (r^2+a^2)\sin^2\theta d\varphi^2
+ {m\over{\rho^2}}(dt + a\sin^2\theta d\varphi)^2 \cr
& \qquad + {{\rho^2}\over{r^2 + a^2 - m}} dr^2
+ \rho^2 d\theta^2 + r^2 \cos^2\theta d\psi^2 + dy^2
}}
where $\rho^2 = r^2 + a^2 \cos^2 \theta$.
This black hole can be thought of as a solution of the six dimensional
low energy action of heterotic string theory.  It is a solution which
has only the metric excited but no gauge fields, antisymmetric tensor,
dilaton, or moduli fields turned on.  {}From it, we will eventually
obtain via string/string duality a charged spinning black hole
solution of the Type II theory in five dimensions.  This black hole
will be a spinning generalization of the solution in \stromvafa.

Before we explain our method for generating the final solution, let us
review some salient features of low--energy actions for the heterotic
and Type II theories in six and five dimensions.  We will work with
the six dimensional actions in the sector with just one abelian gauge
field $F={d}\wedge{A}$ and no moduli.  This gauge field is taken to be
a right--handed\foot{We take the field to be right--handed so that the
extremal configuration is supersymmetric} internal gauge field on the
heterotic side, and a field of Ramond--Ramond origin on the Type II
side.  Our notation is such that heterotic fields are denoted by
primes, six dimensional fields have a subscript $6$ so as to
distinguish them from five dimensional fields, and we use the
conventions of \stromvafa.  We have on the heterotic side, to lowest
order in $\alpha^\prime$ \senreview\
$$
S_{het}(T^4) = \int \, d^6x \sqrt{-g^\prime_6} e^{-2\phi^\prime_6} 
\left[ R^\prime_6 + 4(\partial_\mu \phi^\prime_6)^2 
- {{1}\over{12}} H^{\prime\,2}_{6\,\mu\nu\lambda}
- {{1}\over{4}}F^{\prime\,2}_{6\,\mu\nu}              
\right]
$$
with  $\mu=0 ,\ldots , 5$ and 
$$
H^\prime_{6\,\mu\nu\lambda}= \partial_\mu B^\prime_{6\,\nu\lambda}
- {{1}\over{2}}A^\prime_{6\,\mu} F^\prime_{6\,\nu\lambda} 
+ ({\rm cyclic})
$$
Note that the Chern--Simons terms come from the internal gauge fields.
For Type IIA
$$\eqalign{
S_{IIA}(K3)  &= \int \, d^6x 
\left[ \sqrt{-g_6} 
  \left[ e^{-2\phi_6} 
    \left( R_6 + 4(\partial_\mu \phi_6)^2 
    - {{1}\over{12}}H_{6\,\mu\nu\lambda}^2 
    \right) 
  - {{1}\over{4}}F_{6\,\mu\nu}^2 
  \right]
\right. \cr & 
\left. \qquad\qquad\qquad 
- {{1}\over{4}} \epsilon^{\mu\nu\lambda\rho\alpha\beta}
B_{6\,\mu\nu} F_{6\,\lambda\rho} F_{6\,\alpha\beta}
\right]
}
$$
where
$$
H_{6\,\mu\nu\lambda} = \partial_\mu B_{6\,\nu\lambda} + ({\rm cyclic})
$$
These two actions are related by string/string duality
$$
\eqalign{
\phi_6            & = - \phi_6^\prime \cr
g_{6\,\mu\nu}     & = e^{-2\phi^\prime_6} g^\prime_{6\,\mu\nu} \cr
A_{6\,\mu}        & = A^\prime_{6\,\mu} \cr
H_{6\,\mu\nu\lambda} & = {{1}\over{6}}
\epsilon^\prime_{\mu\nu\lambda\rho\alpha\beta} 
\sqrt{-g^\prime_6} e^{-2\phi^\prime_6} H_6^{\prime\,\rho\alpha\beta} 
}
$$
Using the standard Kaluza--Klein reduction on the $S^1$ with
coordinate $y=x^5$,
$$
\eqalign{
ds_6^2 &= {\tilde{g}}_{\mu\nu} dx^\mu dx^\nu 
+ e^{2\sigma} (dy + V_\mu dx^\mu)^2 \cr
\phi_6 &= \phi + {{1}\over{2}}\sigma \cr
B_{6}  &= {{1}\over{2}} [B_{\mu\nu} - {{1}\over{2}}
(V_\mu {\tilde{B}}_\nu - {\tilde{B}}_\mu V_\nu)]dx^\mu \wedge dx^\nu
+ {\tilde{B}}_\mu dx^\mu \wedge dy
}
$$
(in the sector with $A_y=0$) one finds that
$$
\eqalign{
S_{IIA}(K3\times S^1) &= \int \, d^5x 
\left[ \sqrt{-{\tilde{g}}} 
  \left[ e^{-2\phi} 
    \left({\tilde{R}} + 4(\partial_\mu \phi)^2 
    - {{1}\over{12}} H^{2}_{\mu\nu\lambda} 
    \right.
  \right. 
\right. \cr          & 
\left. 
  \left. 
    \left. \qquad
    - (\partial_\mu \sigma)^2
    - {{1}\over{4}} e^{2\sigma} V^2_{\mu\nu}
    - {{1}\over{4}} e^{-2\sigma} {\tilde{H}}^2_{\mu\nu} 
    \right)
  \right. 
\right. \cr          & 
\left. 
  \left. \qquad
  - {{1}\over{4}} e^{\sigma} F^2_{\mu\nu} 
  \right] 
+ {{1}\over{4}}\epsilon^{\mu\nu\lambda\alpha\beta} 
{\tilde{B}}_\mu F_{\nu\lambda} F_{\alpha\beta} 
\right]
}
$$
where $\tilde H = d\wedge\tilde B$ and
$$
H_{\mu\nu\lambda} = \partial_\mu B_{\nu\lambda} 
- {{1}\over{2}} V_\mu {\tilde{H}}_{\nu\lambda} 
- {{1}\over{2}} {\tilde{B}}_\mu V_{\nu\lambda} + ({\rm cyclic})
$$
In five dimensional Einstein frame, defined by
$g_{\mu\nu}=e^{-4\phi/3}{\tilde{g}}_{\mu\nu}$, we obtain
$$
\eqalign{
S_{IIA}(K3\times S^1) &= \int \, d^5 x 
\left[ \sqrt{-g} 
  \left( R - {{4}\over{3}}(\partial_\mu\phi)^2 
  - (\partial_\mu\sigma)^2 
  \right. 
\right. \cr          & 
\left. 
  \left. \qquad
  - {{1}\over{4}} e^{2\sigma - 4\phi/3} V^2_{\mu\nu} 
  - {{1}\over{4}} e^{-2\sigma - 4\phi/3} {\tilde{H}}^2_{\mu\nu}
  - {{1}\over{4}} e^{8\phi/3} X^2_{\mu\nu} 
  - {{1}\over{4}} e^{\sigma + 2\phi/3} F^2_{\mu\nu} 
  \right) 
\right. \cr          & 
\left.  \qquad
+ {{1}\over{4}} \epsilon^{\sigma\rho\mu\nu\lambda} 
(X_\sigma V_{\rho\mu} {\tilde{H}}_{\nu\lambda} 
+{\tilde{B}}_\sigma F_{\rho\mu} F_{\nu\lambda})
\right]
}
$$ 
where in this action we have Hodge--dualized the three--form $H$ via
$$
\eqalign{
H_{\mu\nu\lambda} 
&= {{1}\over{2}} e^{8\phi/3} \sqrt{-g}
\epsilon_{\sigma\rho\mu\nu\lambda} X^{\sigma\rho} \cr
&= \partial_\mu B_{\nu\lambda} 
- {{1}\over{2}} V_\mu {\tilde{H}}_{\nu\lambda}
- {{1}\over{2}} {\tilde{B}}_\mu V_{\nu\lambda} + {\rm cyclic} 
}
$$

Our method for generating the desired black hole is to use a series of
transformations, namely boosts involving the time $t$ and the circle
coordinate $y$, and string/string duality, as follows.  We begin with
the metric \kerrfived\ as a heterotic solution in six dimensions.  We
apply $O(6,6)$ boosts in the $(t,y)$ directions, following the five
dimensional black hole construction of \horsen.  In their notations,
we use boost parameters $x=\cosh\alpha$, and $\beta=-\alpha$.  The
resulting six--dimensional solution has no $G_{6\,y\mu}$ for $\mu<5$,
but has a $B_6$ and a $\phi_6$.  Next, we notice that since this
boosted solution has no gauge ($A_6$) fields, the field configuration
is a solution for the Type II theory as well.  We then apply
string/string duality to get back a heterotic solution.  Using the
$O(6,6)$ symmetry we can thus apply a final boost, mixing $t$ and the
internal direction involving $A^\prime_6$, with parameter
$z=\cosh\gamma$.  We then apply string/string duality to convert the
heterotic solution to a Type II solution, and lastly we perform the
standard Kaluza--Klein reduction to five dimensions.  The above boost
parameter $z$ is carefully chosen to satisfy $z=2x^2-1$,
i.e. $\gamma=2\alpha$; this choice reduces the five dimensional
dilaton to a constant.  The resulting configuration is a charged
spinning five dimensional black hole with constant dilaton and
constant moduli.

\newsec{The extremal black holes}

Here we will exhibit the extremal limit of the black holes obtained
via the procedure outlined above.  To do this, we take the limit
$x\to\infty$, $a\to 0$, $m\to 0$, such that the quantities
$\mu\equiv m\,x^2$ and $\omega\equiv a\,x$ remain finite, where $m,a$
are the quantities appearing in the metric \kerrfived.  After doing a
coordinate transformation to match with \stromvafa, 
$r^2 \to r^2 + \mu$, we obtain for the extremal metric
$$
\eqalign{ ds_5^{2\,(ext)} 
&= -{\left(1 - { \mu \over{r^2}} \right)^2} 
\left[dt - {{\mu\omega\sin^2\theta}\over{(r^2-\mu)}} d\varphi
+ {{\mu\omega\cos^2\theta} \over{(r^2-\mu)}} d\psi \right]^2 \cr
& \qquad + \left( 1 - {\mu\over{r^2}} \right)^{-2} dr^2
+ r^2(d\theta^2 + \sin^2\theta d\varphi^2 + \cos^2\theta d\psi^2) 
\cr
}
$$
while for the Ramond--Ramond gauge field the nonvanishing components
are
$$
\eqalign{
A_t^{(ext)} &= {\sqrt{2}\over\lambda}{{\mu}\over{r^2}}\cr
A_\varphi^{(ext)} &=  {\sqrt{2}\over\lambda}
                 {{\omega\mu\sin^2\theta}\over{r^2}} \cr 
A_\psi^{(ext)}    &= -{\sqrt{2}\over\lambda}
                 {{\omega\mu\cos^2\theta}\over{r^2}}
}
$$
and for the winding gauge field we have
$$
\tilde B^{(ext)} = {{\lambda^3}\over{\sqrt{2}}} A^{(ext)}
$$
While the result of our solution generating procedure yields
$\phi = 0 = \sigma$, we have shifted these scalars by a constant to
$$
e^{\sigma+2\phi/3}=\lambda^2
$$
which introduces the scaling of the gauge fields by $\lambda$
given above \stromvafa. The
above fields are the only ones excited in this black hole background.
Notice that when we take $\omega\to 0$, we recover the solution of
\stromvafa.

{}From the asymptotic metric we obtain for the angular momentum, in
the independent planes defined by $\varphi,\psi$,
$$\eqalign{
J_1 &\equiv J_{\varphi} = +{\pi\over{4}} \mu \omega \cr
J_2 &\equiv J_{\psi}    = -{\pi\over{4}} \mu \omega
}$$
and for the mass we find
$$
M_{ADM} = {{3\pi\mu}\over{4}}
$$
while the charges are\foot{The sphere $S^3$ is at infinity, so we
can ignore the effects of the Chern--Simons terms}
$$
\eqalign{
Q_H &\equiv {{1}\over{4\pi^2}} \int_{S^3} 
{}^\star e^{-2\sigma-4\phi/3}\tilde H = \mu/\lambda^2 \cr
Q_F &\equiv {{1}\over{16\pi}} \int_{S^3}
{}^\star e^{\sigma+2\phi/3} F = -{{\pi}\over{2\sqrt{2}}}\mu\,\lambda
}
$$ 
Note that this black hole, although a solution of the low--energy
string theory equations, is not a solution of the Einstein--Maxwell
equations in five dimensions.  In the spinning configuration, the
magnetic dipole field combines with the electric monopole field so
that the Chern--Simons contributions to the equations of motion
are nontrivial.  (The terms we are referring to are of course distinct
from the usual ${\cal O}(\alpha^\prime)$ Chern--Simons terms already
appearing in the ten dimensional heterotic string theory; we are not
including those terms here, as we are doing our analysis to lowest
order in $\alpha^\prime$.)

Let us now obtain the entropy of this extremal spinning black hole.
In the above coordinates, the horizon is at $r=r_0=\sqrt{\mu}$, and
its entropy is found to be ($|J_1|=|J_2|\equiv J$)
\eqn\sbh{\eqalign{
S_{BH} &= {{1}\over{2}} \pi^2 \mu\sqrt{\mu - \omega^2} \cr
       &= 2\pi\sqrt{{{Q_H\,Q_F^2}\over{2}}-J^2}
}}
Note that both of these expressions are independent of $\lambda$.

We find it satisfying that these extremal rotating charged black holes
have a finite--area horizon, and also that the angular momentum is
bounded above: $J^2_{max}=Q_H\,Q_F^2/2$ (in going beyond this limit,
closed timelike curves develop).

\newsec{Spins of the BPS D--branes}

Let us recall the nature of the D--brane states that are responsible
for the degeneracy of the extremal black holes that we are considering
\stromvafa.  We will consider compactification of type IIB on 
$K3\times S^1$ down to five dimensions\foot{ The arguments for
compactification on $T^4\times S^1$ are (essentially) identical with
the replacement of $T^4$ for $K3$ in the following discussions.  Only
the dimension of the manifold enters in the asymptotic growth below.}.
Consider those D--brane states which are wrapped around $S^1$ and
partially wrapped around $K3$. Let $Q_F$ denote the charge of the
D--brane on the $K3$ part. $Q_F$ can be viewed as an element of the
$K3$ homology $H_*(K3,{\bf Z})$ which is identified with how the
internal part of the D--brane wraps around $K3$.  Note that the dot
product $Q_F\cdot Q_F$ is the same as the intersection of cycles in
the $K3$ homology.  In the presence of D--branes, we get an effective
field theory which lives on the D--brane worldvolume.  If we take the
size of $S^1$ to be much larger than that of the $K3$, then the
D--brane effective field theory will be a theory on ${\bf R}\times
S^1$, where the ${\bf R}$ is the time coordinate.

Based on string dualities and observations in \vafawitten\ it was
conjectured \vafagas\ that this theory is a sigma model on
$({{1}\over{2}}Q_F\cdot Q_F+1)$ symmetric product of $K3$'s, i.e. on
$$
M={\rm Sym}^{{1\over 2}Q_F^2+1}(K3)
$$
This conjecture has been checked in essentially all cases, at least up
to $T$-duality, and found to be true \bersadvafa, \vafainst.
Actually, as noted in \vafawitten, \vafagas, the light-cone helicity
of the six dimensional theory, which we identify with the spatial
$O(4)$ holonomy, can also be read off, as follows.  Let $J_1$ and
$J_2$ be the two holonomies in $O(4)$.  Note that the sigma model on
$M$ is conformal (as it is hyperkahler) and it will give rise to two
$U(1)$'s from the $N=2$ superconformal algebras: one left-- and one
right--moving.  In fact, there will be an $N=4$ superconformal algebra
in our case, with the $SU(2)_L\times SU(2)_R$ action to be identified
with our $O(4)$, but we will only need the $U(1)_L\times U(1)_R$
subgroup of it here.  Let us denote the $U(1)_L\times U(1)_R$ charges
of states by $(F_L,F_R)$.  Then
$$
\eqalign{
J_1 &= {1\over 2}(F_L+F_R) \cr
J_2 &= {1\over 2}(F_L-F_R)
}
$$
Consider, for example, the case $Q_F=0$.  Noting that the ground
states of the sigma model are identified with the $K3$ cohomology, and
that $F_L$ and $F_R$ for the ground states run over the values
$\{-1,0,1\}$, we learn that we have the $(J_1,J_2)$ spectrum
consisting of 20 states with (0,0), two states with $(\pm 1,0)$, and
two states with $(0,\pm 1)$.  These we recognize as the light-cone
oscillator quantum numbers of bosonic strings in 6 dimensions.

The D--brane BPS states considered in \stromvafa\ correspond to
Ramond-Ramond states of this sigma model which are right--moving
ground states, and left--moving states of level $n=Q_H$.  Recall that
there is a bound for the $F_L$ and $F_R$ with respect to $L_0$ and
$\overline{L_0}$ \bouchfriedkent.  This is easily seen by bosonizing
the $U(1)$ currents: let $J_L=\sqrt{\hat c} \partial \phi$.  A state
with charge $F_L$ will then be represented by an operator
$$
{\rm exp}({i F_L\phi \over \sqrt{\hat c}})\cdot \Phi
$$
where $\Phi$ is an operator from the rest of the conformal field
theory which can be made of the oscillator factors of the $U(1)$
current, but not the momentum modes, plus any other state in the
theory.  The same story repeats for $F_R$.  In particular, note that
since the dimensions of $\Phi$ are positive, the dimensions of the
operators are restricted by
$$
L_0\geq {F_L^2\over 2\hat c} \qquad \overline{L_0}
\geq {F_R^2\over 2\hat c}
$$
where $\hat c$ is the complex dimension of the manifold.  In our case,
$\hat c =Q_F^2+2$.

We are interested in doing a count in the regime where $Q_F$ is large
but held fixed.  Moreover, we take $Q_H$ to be
arbitrarily large.  We are also interested in a region with
$|J_1|,|J_2|>>1$. 
Let us consider the case where the system is the right-moving ground
state with fixed $F_R$. Then we can consider arbitrarily
large values of $F_L$ to make both $J_1$ and $J_2$ large with the
same sign\foot{To consider $J_1$ and $J_2$ with the opposite sign,
the entropy would have come from
the right-movers and we would be considering large values of $F_R$.}.
Since the entropy comes from the left-moving Hilbert space, we have to
estimate how many left--mover states are still available if we fix
$F_L$. Considering a regime\foot{It may be that our final results
are valid beyond this regime of charges. Further note that the given
regime does not exclude the possibility that the ratio of $F_L^2/2\hat c$
to $Q_H$ is only slightly less than one.}
where $(Q_H-F_L^2/2\hat c)>>1$ as well as $Q_H/Q_F^2>>1$,
the answer is supplied by the bosonization discussed above.
Since the total eigenvalue is $L_0=n=Q_H$, and we have used up
${F_L^2\over 2\hat c}={F_L^2\over 2Q_F^2+4}$ for the states we are
interested in, the $L_0$ eigenvalue of the extra operator $\Phi$ is
given by
$$
L_0(\Phi)=\tilde n=n-{F_L^2\over 2\hat c}=Q_H-{F_L^2\over 2Q_F^2+4}
$$
Since the oscillatory states make the maximum contribution to
degeneracy of string states, we learn that effectively we can take
$\tilde n$ as the available oscillator number.  Therefore, we get a
degeneracy growth of ($\hat c=c/3$, $F_L=J_1+J_2$)
$$
\eqalign{
d\sim{\rm exp}(2 \pi \sqrt{\tilde n \hat c\over 2})&=
{\rm exp}\left(2 \pi \sqrt{(Q_H-{F_L^2/4\over 
{1\over 2}Q_F^2+1})({1\over 2}Q_F^2+1)}\right) \cr
&\sim {\rm exp}\left(2\pi \sqrt{Q_H({1\over 2}Q_F^2+1)-{1\over 4}
(|J_1|+|J_2|)^2}\right)
}
$$
We have used absolute value signs for $J_i$ in order to write the
final answer in its most general form, independently of whether
or not
$J_1$ and $J_2$ have the same sign.  Then the entropy is 
\eqn\smicro{
S_{micro} \sim 2\pi \sqrt{Q_H({1\over 2}Q_F^2+1)-{1\over 4}
(|J_1|+|J_2|)^2}
}
Taking $|J_1|=|J_2|=J$, we see that this formula agrees with what we
found for the entropy \sbh\ of the spinning black hole.  Note that
this computation also sharpens the computation in \stromvafa\ where,
in principle, one should have counted only the spin--zero D--branes to
make the comparison with the non--spinning black hole.

\newsec{Comments and Acknowledgements}

In this work we have found spinning black hole solutions in five
dimensions, which we believe are new, whose entropy agrees precisely
with the result obtained by using D--brane technology.  It is
satisfying that the two methods give the same result, and we regard
this as additional evidence for the D--brane picture of \polchinski.

One would also like to consider the case where the two angular
momenta, $J_1$ and $J_2$, are not equal. With respect to our
D-brane calculation, we note that in the right-moving ground state
$|F_R|$ is bounded as \bouchfriedkent
$$
|F_R|=|J_1-J_2|\le{{\hat c}\over 2}={1\over 2}Q_F^2 +1
$$
Hence the difference between the spins can not be arbitrarily large.
By combining this bound with the previously noted relation
$Q_H/Q_F^2>>1$, one can demonstrate that our calculations are
valid for $|F_R|/|F_L|=|J_1-J_2|/|J_1+J_2|<<1$.
Hence one should not expect to see a difference in the spins at the
macroscopic level of the black hole computations. In fact, one can
construct a nonextremal black hole analogous to that presented here
in which the two angular momenta are independent. However, one finds
demanding that the extremal or supersymmetric limit be nonsingular
requires setting $|J_1|=|J_2|$ \nonext. Hence the D-brane and
black hole results are in perfect agreement on this
aspect of the calculation, as well.

Lastly, we would like to point out that in the degeneracy formula
there are power corrections which give rise to logarithmic corrections
to the entropy \smicro.  These should correspond to logarithmic
corrections to the entropy on the black hole side, and it would be
interesting to investigate whether these do in fact arise as one--loop
corrections in the black hole geometry.

Work on the nonextremal versions of the black holes exhibited here is
in progress.  It appears that the results of \callanmalda, \horstrom\
can be extended to the nonextremal spinning case \nonext.

\bigskip

We would like to thank A. Strominger and E. Witten for communications.

\medskip

The research of JCB and RCM was supported by NSERC of Canada and Fonds
FCAR du Qu\'ebec; that of AWP by NSF grant PHY-90-21984; and that of
CV by NSF grant PHY-92-18167.

\listrefs
\bye